\documentclass[preprint,superscriptaddress,preprintnumbers,amsmath,amssymb,prd]{revtex4}
\usepackage{graphicx}

\begin{document}
\def\Xint#1{\mathchoice
   {\XXint\displaystyle\textstyle{#1}}%
   {\XXint\textstyle\scriptstyle{#1}}%
   {\XXint\scriptstyle\scriptscriptstyle{#1}}%
   {\XXint\scriptscriptstyle\scriptscriptstyle{#1}}%
   \!\int}
\def\XXint#1#2#3{{\setbox0=\hbox{$#1{#2#3}{\int}$}
     \vcenter{\hbox{$#2#3$}}\kern-.5\wd0}}
\def\ddashint{\Xint=}
\def\dashint{\Xint-}

\thispagestyle{empty}

\title{Kramers-Kronig relations and causality conditions for graphene
in the framework of the Dirac model}

\author{G.  L.  Klimchitskaya}
\affiliation{Central Astronomical Observatory at Pulkovo of the
Russian Academy of Sciences, Saint Petersburg,
196140, Russia}
\affiliation{Institute of Physics, Nanotechnology and
Telecommunications, Peter the Great Saint Petersburg
Polytechnic University, Saint Petersburg, 195251, Russia}

\author{
V.~M.~Mostepanenko}
\affiliation{Central Astronomical Observatory at Pulkovo of the
Russian Academy of Sciences, Saint Petersburg,
196140, Russia}
\affiliation{Institute of Physics, Nanotechnology and
Telecommunications, Peter the Great Saint Petersburg
Polytechnic University, Saint Petersburg, 195251, Russia}
\affiliation{Kazan Federal University, Kazan, 420008, Russia}

\begin{abstract}
We analyze the concept of causality for the conductivity of
graphene described by the Dirac model. It is recalled that
the condition of causality leads to the analyticity of
conductivity in the upper half-plane of complex frequencies
and to the standard symmetry properties for its real and
imaginary parts. This results in the Kramers-Kronig relations,
which explicit form depends on whether the conductivity has
no pole at zero frequency (as in the case of zero temperature
when the band gap of graphene is larger than twice the
chemical potential) or it has a pole (as in
all other cases, specifically, at nonzero temperature).
Through the direct analytic calculation it is shown that
the real and imaginary parts of graphene conductivity,
found recently on the basis of first principles of thermal
quantum field theory using the polarization tensor in
(2+1)-dimensional space-time, satisfy the Kramers-Kronig
relations precisely. In so doing, the values of two integrals
in the commonly used tables, which are also important for a
wider area of dispersion relations in quantum field theory
and elementary particle physics, are corrected. The obtained
results are not of only fundamental theoretical character,
but can be used as a guideline in testing the validity of
different phenomenological approaches and for the
interpretation of experimental data.
\end{abstract}

\maketitle

\section{INTRODUCTION}

Considerable recent attention has been focused on graphene,
which is a two-dimensional sheet of carbon atoms packed in
a hexagonal lattice \cite{1,2}. This unique material is
interesting not only for condensed matter physics due to its
unusual electrical and mechanical properties, but for
quantum field theory as well. The point is that the electronic
excitations in graphene are either massless or very light. At
energies below a few eV they possess the linear dispersion
relation and obey (2+1)-dimensional Dirac equation where the
speed of light $c$ is replaced with the Fermi velocity
$v_{F}\approx c$/300 \cite{1,2,3}. Thus, graphene makes possible
testing many predicted effects of quantum field theory and
quantum electrodynamics which are not experimentally
feasible with much heavier ordinary electrons. Among other
effects one could mention the Klein paradox \cite{4}, the
creation of particle-antiparticle pairs from vacuum in a
static \cite{5,6} and time-dependent \cite{7,8} electric
field, and the relativistic quantum Hall effect in a
strong magnetic field \cite{9}.

Graphene is also unique in that its response to external
electromagnetic field and quantum fluctuations, described
by the polarization tensor in (2+1)-dimensional space-time,
can be found in an explicit form on the basis of first
principles of thermal quantum field theory. Although some
special cases have been considered previously (see, e.g.,
Ref.~\cite{10} and literature therein), the complete
expression for the polarization tensor of graphene in the
one-loop approximation has been derived at zero
temperature in Ref.~\cite{11} and at any nonzero
temperature in Ref.~\cite{12}, where the area of
application was limited to the pure imaginary Matsubara
frequencies. In doing so both cases of zero and nonzero
width of the gap $\Delta$ between the energy bands
(i.e., of gapless and gapped
graphene) and chemical potential $\mu$ were considered.
The results of Refs.~\cite{11,12} have been extensively
used when investigating the Casimir and Casimir-Polder
forces in graphene systems
\cite{13,14,15,16,17,18,19,20,21,22,23}
(some other, more phenomenological, approaches used for
this purpose are the density-density correlation functions,
models of the response functions of graphene by
Lorenz-type oscillators, and the Kubo formalism
\cite{24,25,26,27,28,29,30,31,32,33,34,35}).

A more universal representation for the polarization tensor
of graphene at nonzero temperature was derived in
Ref.~\cite{36}. Unlike Ref.~\cite{12}, the polarization tensor
of Ref.~\cite{36} allows an analytic continuation to the
entire plane of complex frequencies including the real
frequency axis. At the pure imaginary Matsubara frequencies
both representations take the same values. The novel
representation was applied in investigations of the
Casimir force \cite{37,38,39,40,40a} and, after a continuation
to the real frequency axis, for better understanding of
the reflectances of graphene and graphene-coated plates
\cite{41,42,43}. In Ref.~\cite{44} the polarization tensor
of Ref.~\cite{36} was generalized for the case of doped
graphene with a nonzero chemical potential. This
generalization was used \cite{45} to investigate an
impact of nonzero band gap and chemical
potential on the thermal effect in the Casimir force.

One of the most important characteristics of graphene is
its electrical conductivity. This quantity possesses
many surprising properties connected with an existence
of the so-called \textit{universal} conductivity
$\sigma_{0}=e^{2}/(4\hbar)$ expressed via the fundamental
constants, electron charge $e$ and Planck constant $\hbar$.
For a pure graphene, having the zero band gap
and no doping, the conductivity is equal to $\sigma_{0}$
in the limit of zero temperature. This result might be
considered as paradoxical if to take into account that
with vanishing temperature the concentration of charge
carriers in pure graphene goes to zero and there is no
scattering and no dissipation processes.

The conductivity of graphene was extensively
investigated by many authors using the current-current
correlation functions, the Kubo formalism, the Boltzmann
transport theory, and the two-dimensional Drude model
(see the review papers \cite{46,47,48} and references
therein). Some of the results obtained
 employ simple intuitive models,
phenomenological approaches of a limited application
area and even do not agree with each other.
To overcome these troubles, the conductivity of
graphene at any temperature was investigated on the basis
of first principles of quantum electrodynamics using the
polarization tensor of Refs.~\cite{36,44} analytically
continued to the real frequency axis. In
Refs.~\cite{49,50} the cases of pure and gapped
graphene were considered, respectively, and
in Ref.~\cite{51} of both gapped and doped graphene
characterized by nonzero band gap $\Delta$ and
chemical potential $\mu$. The real and imaginary parts of
graphene conductivity have been found in an explicit form. It was
shown that the major contribution to the conductivity of
graphene calculated in the framework of Dirac model is
local, whereas the nonlocal corrections are negligibly
small.

In this paper, we consider the problem of causality in
the response of graphene to electric field. The demand
of causality leads to some constraints on the local
conductivity of graphene. Specifically, it should be an
analytic function in the upper half-plane of complex
frequencies and satisfy certain symmetry conditions.
These result in the Kramers-Kronig relations for the
real and imaginary parts of the conductivity of
graphene. Until the present time the Kramers-Kronig
relations for graphene were discussed only using some
approximate, phenomenological approaches leading to
incomplete and even contradictory results (see, e.g.,
Refs.~\cite{52,52a,53,54}).
Thus, the form of Kramers-Kronig relations used in
Refs.~\cite{52,52a,53} does not take into account that the imaginary
part of the conductivity of graphene has a pole at zero frequency.
Furthermore, Ref. \cite{52} arrives to the Kramers-Kronig relation
expressing the real part of graphene conductivity via its imaginary
part, but fails in obtaining a similar relation with interchanged
real and imaginary parts. An existence of the universal conductivity
of graphene $\sigma_0$ is not taken into account. Moreover,
Ref.~\cite{54} admits that under some conditions the Kramers-Kronig
relations for graphene do not hold. When it is considered that
these relations are not only of fundamental theoretical character,
but are used for interpretation of the measurement data (see, e.g.,
Ref.~\cite{53}), it is of prime importance to conclusively find out
their specific form for graphene and directly prove their validity.

Below we establish an explicit form of the Kramers-Kronig
relations for graphene and demonstrate
that the real and imaginary parts of its conductivity,
 found independently on the basis of
first principles of quantum electrodynamics, satisfy
these relations precisely. Depending on
temperature and a relationship between the band gap
$\Delta$ and chemical potential $\mu$, an additional
pole term in the Kramers-Kronig relations may arise
as it holds in a familiar case of metals \cite{55}.
The obtained relations take proper account for the universal
conductivity of graphene $\sigma_{0}$. In fact, there
is no reason that the Kramers-Kronig relations were not
satisfied for the conductivity determined from the first
principles. The obtained results, however, are physically
meaningful because they establish the specific form of the
Kramers-Kronig relations for so unusual material as graphene
and, by performing the direct verification of these relations,
confirm the expressions for its conductivity found recently
in Ref.~\cite{51} using the polarization tensor.
In the course of our calculations, the values of two
integrals, indicated incorrectly in the most
comprehensive and widely used table of integrals
\cite{56}, have been corrected. These integrals might
be also useful in a wider context of dispersion
relations for the scattering amplitudes in quantum
field theory and physics of elementary particles.

The paper is organized as follows. In Sec.~II, the
brief summary for the polarization tensor, conductivity
of graphene and causality conditions is presented.
Section III contains the proof of the Kramers-Kronig
relations for the conductivity of graphene at zero
temperature. The validity of the Kramers-Kronig
relations at nonzero temperature is demonstrated in
Sec.~IV. In Sec.~V, the reader will find our
conclusions and a discussion. Appendices A and B
contain some details of several mathematical
derivations.

\section{Polarization tensor, conductivity of graphene and
causality conditions}

The polarization tensor of graphene in the one-loop approximation
in the momentum representation is defined according to Refs.~\cite{57,58}
with the following differences. We consider the (2+1)-dimensional
space-time. In the free Dirac equation the speed of light $c$ is replaced with
the Fermi velocity $v_F\approx c/300$ although an interaction with the
electromagnetic field is governed, as usually, by the
coupling constant $e/c$.
In addition, one should take into account that we consider the polarization
tensor at nonzero temperature $T$. Because of this, according to the
Matsubara formalism, an integration over the zeroth component $q_0$ of the
wave vector $q^{\mu}$ of a loop electronic excitation should be replaced
with a summation over the pure imaginary fermionic Matsubara frequencies
\begin{equation}
cq_{0n}=2\pi i\left(n+\frac{1}{2}\right)\frac{k_BT}{\hbar},
\label{eq1}
\end{equation}
\noindent
where $k_B$ is the Boltzmann constant and $n=0,\,\pm 1,\,\pm 2,\,\ldots\,$.
Finally it is necessary to replace the zeroth component $k_0$ of the wave
vector $k^{\mu}$ of an external photon in the argument of the polarization
tensor with the pure imaginary bosonic Matsubara frequencies
\begin{equation}
i\xi_l=ck_{0l}=2\pi i l\frac{k_BT}{\hbar},
\label{eq2}
\end{equation}
\noindent
where  $l=0,\,\pm 1,\,\pm 2,\,\ldots\,$.

As a result, the polarization tensor takes the form \cite{11,12,36,57,58}
\begin{eqnarray}
&&
\Pi^{\mu\nu}(i\xi_l,\mbox{\boldmath$k$})=-8\pi\alpha k_BT
\sum_{n=-\infty}^{\infty}\left(n+\frac{1}{2}\right)
\label{eq3} \\
&&
\times\!\!\! \int\!\!
\frac{d\mbox{\boldmath$q$}}{(2\pi)^2}{\rm tr}
\frac{1}{i\tilde{\gamma}^{\mu}q_{\mu}-\Delta/(2\hbar)}\tilde{\gamma}^{\mu}
\frac{1}{i\tilde{\gamma}^{\mu}q_{\mu}-i\tilde{\gamma}^{\mu}k_{\mu}-
\Delta/(2\hbar)}.
\nonumber
\end{eqnarray}
\noindent
Here, $\alpha=e^2/(\hbar c)\approx 1/137$ is the fine structure constant,
$q^{\mu}=(q_{0n},q^1,q^2)$, $k^{\mu}=(k_{0l},k^1,k^2)$, $\mu=0,\,1,\,2$,
$\mbox{\boldmath$k$}=(k^1,k^2)$,
$\tilde{\gamma}^{\mu}=\eta_{\nu}^{\,\mu}\gamma^{\nu}$ where
$\eta_{\nu}^{\,\mu}={\rm diag}(c,v_F,v_F)$ and $\gamma^{\nu}$ are the
Dirac matrices. Note also that the numerical factor
on the right-hand side of Eq.~(\ref{eq8}) takes into account four fermion
species for graphene \cite{1,2,3}.

The polarization tensor (\ref{eq3}) was calculated
over the entire axis of imaginary frequencies in Ref.~\cite{36},
analytically continued
to the real frequency axis and used for different purposes in
Refs.~\cite{36,41,42,43,49,50}. In Ref.~\cite{44} this tensor was
generalized for the case of graphene with nonzero chemical potential $\mu$
[this is reached by the replacement $q_{0n}\to q_{0n}+\mu/(\hbar c)$]
and analytically continued to the real frequency axis in Ref.~\cite{51}.
The longitudinal (in-plane of graphene) and transverse (out-of-plane)
electrical conductivities are expressed via the polarization tensor as
\cite{22,49,50,51}
\begin{eqnarray}
&&
\sigma_{\|}(\omega,k,T)=-i\frac{\omega}{4\pi\hbar k^2}\Pi_{00}(\omega,k,T),
\nonumber \\
&&
\sigma_{\bot}(\omega,k,T)=i\frac{c^2}{4\pi\hbar\omega k^2}\Pi(\omega,k,T),
\label{eq4}
\end{eqnarray}
\noindent
where
\begin{equation}
\Pi(\omega,k,T)=k^2{\rm tr}\Pi_{\mu\nu}(\omega,k,T)+
\left(\frac{\omega^2}{c^2}-k^2\right)\Pi_{00}(\omega,k,T)
\label{eq5}
\end{equation}
\noindent
and $k=|\mbox{\boldmath$k$}|$. The conductivities of graphene are the
complex quantities as well as the polarization tensor along the real
frequency axis.

Calculations show that the major contributions to both the real and imaginary
parts of $\sigma_{\|(\bot)}$ are given in the local limit $k=0$,
whereas the nonlocal corrections are of the order of $(v_F/c)^2\sim 10^{-5}$.
In the local limit one has
\begin{equation}
\sigma(\omega,T)\equiv\sigma_{\|}(\omega,0,T)=
\sigma_{\bot}(\omega,0,T).
\label{eq6}
\end{equation}
\noindent
Note that the quantities $\Pi_{00}$ and $\Pi$ in Eq.~(\ref{eq4})
go to zero as $k^{2}$ when $k$ goes to zero, whereas ${\rm tr}\Pi_{\mu\nu}$
goes to a nonzero constant. Expanding all these quantities up to the
first power in the parameter $(v_{F}k/\omega)^{2} < (v_{F}/c)^{2}$
and using Eqs.~(28),~(40), and (43) in Ref.~\cite{51}, one obtains that
in this perturbation order
\begin{equation}
\Pi(\omega,k,T)=-\frac{\omega^2}{c^2}\Pi_{00}(\omega,k,T).
\label{eq6a}
\end{equation}
\noindent
Taking into account Eq.~(\ref{eq4}), it is seen that  Eq.~(\ref{eq6a}) is in 
agreement with Eq.~(\ref{eq6}).

The explicit expressions for the quantity $\sigma(\omega,T)$
in the most general case of graphene with nonzero $\Delta$ and $\mu$
have been derived from Eq.~(\ref{eq4}) in Ref.~\cite{51}.
It is convenient to present the local conductivity of graphene
(\ref{eq6}) as the sum of two contributions
\begin{equation}
\sigma(\omega,T)=\sigma^{(0)}(\omega)+\sigma^{(1)}(\omega,T).
\label{eq7}
\end{equation}
\noindent
The quantity $\sigma^{(0)}$ on the right-hand side of this equation
is the contribution to the conductivity which does not depend on
$T$ and $\mu$. It is given by \cite{51}
\begin{eqnarray}
&&
{\rm Re}\sigma^{(0)}(\omega)=\sigma_0\theta(\hbar\omega-\Delta)
\frac{(\hbar\omega)^2+\Delta^2}{(\hbar\omega)^2},
\label{eq8} \\
&&
{\rm Im}\sigma^{(0)}(\omega)=\frac{\sigma_0}{\pi}\left[
\frac{2\Delta}{\hbar\omega}-
\frac{(\hbar\omega)^2+\Delta^2}{(\hbar\omega)^2}
\ln\left|\frac{\hbar\omega+\Delta}{\hbar\omega-\Delta}\right|
\right],
\nonumber
\end{eqnarray}
\noindent
where $\Delta$ is the width of the gap in Dirac's spectrum and
$\theta(x)$ is the step function equal to unity for $x\geq 0$ and zero
for $x<0$. Note that under the condition $\Delta>2\mu$
the quantity $\sigma^{(0)}(\omega)$ defined in Eq.~(\ref{eq8}) has
the physical meaning of the total conductivity of graphene at zero
temperature $\sigma(\omega,0)$. This means that under the condition
$\Delta>2\mu$ it holds \cite{51}
\begin{equation}
\sigma^{(1)}(\omega,0)=\lim_{T\to 0}\sigma^{(1)}(\omega,T)=0.
\label{eq8a}
\end{equation}
\noindent
Thus, if $\Delta>2\mu$ the conductivity $\sigma(\omega,0)$ does not
depend on $\mu$ (even if $\mu$ is not equal to zero but is smaller
than $\Delta/2$) and ${\rm Re}\sigma^{(0)}$ vanishes if
$\hbar\omega<\Delta$. The conductivity of graphene at $T=0$ and
$\Delta<2\mu$ is considered in Sec.~III.

The quantity $\sigma^{(1)}$ on the right-hand side of Eq.~(\ref{eq7})
depends on $T$, $\Delta$ and $\mu$. It can be represented in the
form \cite{51}
\begin{eqnarray}
&&
{\rm Re}\sigma^{(1)}(\omega,T)=-\sigma_0\theta(\hbar\omega-\Delta)
\frac{(\hbar\omega)^2+\Delta^2}{(\hbar\omega)^2}F(\omega,T),
\label{eq9} \\
&&
{\rm Im}\sigma^{(1)}(\omega,T)=\frac{2\sigma_0}{\pi}
\int_{\frac{\Delta}{\hbar\omega}}^{\infty}\!\!\!\!dt\left[1+
\frac{(\hbar\omega)^2+\Delta^2}{(\hbar\omega)^2}
\frac{1}{t^2-1}\right]F(\omega t,T),
\nonumber
\end{eqnarray}
\noindent
where the function $F(x,T)$ is defined as
\begin{equation}
F(x,T)=\sum_{\kappa=\pm 1}\left[
\exp\left(\frac{\hbar x+2\kappa\mu}{2k_BT}\right)+1\right]^{-1}.
\label{eq9a}
\end{equation}

It is convenient to introduce the new integration variable
$v=\hbar\omega t/\Delta$ in the second line of Eq.~(\ref{eq9})
which takes the form
\begin{equation}
{\rm Im}\sigma^{(1)}(\omega,T)=\frac{2\sigma_0}{\pi}
\frac{\Delta}{\hbar\omega}
\int_{1}^{\infty}\!\!\!\!\!dv\left[1+
\frac{\Delta^2+(\hbar\omega)^2}{(v\Delta)^2-(\hbar\omega)^2}
\right]F\left(\frac{v\Delta}{\hbar},T\right).
\label{eq9b}
\end{equation}
\noindent
This expression has the pole term $C(T)/\omega$ at $\omega=0$, where
\begin{equation}
C(T)=\frac{2\sigma_0}{\pi}
\frac{\Delta}{\hbar}
\int_{1}^{\infty}\!\!\!dv
\frac{v^2+1}{v^2}F\left(\frac{v\Delta}{\hbar},T\right).
\label{eq10}
\end{equation}

Now we separate the pole term in the imaginary part of conductivity by adding and subtracting the quantity $C(T)/\omega$ on the right-hand side of Eq.~(\ref{eq9b}).
Leaving the first expression in Eq.~(\ref{eq9}) unchanged, both
${\rm Re}\sigma^{(1)}$ and ${\rm Im}\sigma^{(1)}$ can be rewritten as
\begin{eqnarray}
&&
{\rm Re}\sigma^{(1)}(\omega,T)=-\sigma_0\theta(\hbar\omega-\Delta)
\frac{(\hbar\omega)^2+\Delta^2}{(\hbar\omega)^2}F(\omega,T),
\nonumber \\
&&
{\rm Im}\sigma^{(1)}(\omega,T)=\frac{C(T)}{\omega}
\label{eq10a} \\
&&~~~~~
+\frac{2\sigma_0}{\pi}\hbar\omega\Delta
\int_{1}^{\infty}\!\!\!\!\!dv
\frac{v^2+1}{v^2[(v\Delta)^2-(\hbar\omega)^2]}
F\left(\frac{v\Delta}{\hbar},T\right).
\nonumber
\end{eqnarray}

Now we discuss the requirements of causality imposed on the conductivity
$\sigma(\omega,T)$ and its constituents $\sigma^{(0)}(\omega)$ and
$\sigma^{(1)}(\omega,T)$.
According to the principle of causality, the electric current density
$\mbox{\boldmath$j$}(t)$ must not depend on the values of electric field
$\mbox{\boldmath$E$}(t)$ at times greater than $t$, i.e.,

\begin{equation}
\mbox{\boldmath$j$}(t,T)=\int_{0}^{\infty}\sigma(\tau,T)
\mbox{\boldmath$E$}(t-\tau)d\tau.
\label{eq11}
\end{equation}
\noindent
Multiplying both sides of this equation by $e^{i\omega t}$ and integrating with
respect to $t$ from $-\infty$ to $\infty$, we obtain an equation for the Fourier
images of the field and current density
\begin{equation}
\mbox{\boldmath$j$}(\omega,T)=\sigma(\omega,T)
\mbox{\boldmath$E$}(\omega),
\label{eq12}
\end{equation}
\noindent
where
\begin{equation}
\sigma(\omega,T)=\int_{0}^{\infty}\sigma(\tau,T)
e^{i\omega\tau}d\tau.
\label{eq13}
\end{equation}

Repeating the well known reasoning contained in Ref.~\cite{55} for the case
of frequency-dependent dielectric permittivity, it is easy to find the
analytic properties of $\sigma(\omega,T)$ in the plane of complex frequencies
and the symmetry properties of its real and imaginary parts.
Specifically, from Eq.~(\ref{eq13}) it follows that in the upper half-plane
(${\rm Im}\omega>0$) $\sigma(\omega,T)$ is an analytic function with no
singularities. The real and imaginary parts of $\sigma(\omega,T)$  are the
even and odd functions of real frequency, respectively. From Eq.~(\ref{eq13})
it is seen also that for the complex $\omega$ it holds
$\sigma(-\omega^{\ast},T)=\sigma^{\ast}(\omega,T)$.
Then at the pure imaginary frequencies $\sigma(\omega,T)$ takes the real
values. Equation (\ref{eq11}) is also valid for the contributions
$\sigma^{(0)}(\tau)$ and $\sigma^{(1)}(\tau,T)$ to the conductivity
$\sigma(\tau,T)$ [with the corresponding contributions to the total current
$\mbox{\boldmath$j$}^{(0)}(t)$ and  $\mbox{\boldmath$j$}^{(1)}(t,T)$
on the left-hand side]. From this it follows that all the above properties
of $\sigma(\omega,T)$ are inherent also in $\sigma^{(0)}(\omega)$ and
$\sigma^{(1)}(\omega,T)$.

Note that the explicit expressions (\ref{eq8}) and (\ref{eq10a}) may appear in disagreement with the formulated above general properties of conductivity following from
the causality condition (\ref{eq11}). The point is that it may exist several
equivalent representations for some quantity along the positive frequency axis,
but only one of them allows immediate analytic continuation to the entire plane
of complex frequencies. Equations (\ref{eq8}) and (\ref{eq10a}) are written in
the form which is most convenient for applications only at the real, positive
frequencies and can be easily compared with the results obtained using various
approximate and phenomenological approaches (see below). These equations,
however, can be identically rewritten in the form where the analytic continuation
from the real, positive frequency axis to the entire complex
frequency plane is achieved
by simply putting frequency $\omega$ complex. For example, Eq.~(\ref{eq8})
can be rewritten in the form
\begin{equation}
\sigma^{(0)}(\omega)=i\frac{2\sigma_0}{\pi}\left[
\frac{\Delta}{\hbar\omega}-
\frac{(\hbar\omega)^2+\Delta^2}{(\hbar\omega)^2}
{\rm arctanh}\frac{\hbar\omega}{\Delta}\right],
\label{eq13a}
\end{equation}
\noindent
where all the above properties are evidently satisfied.
An equivalence of Eqs.~(\ref{eq8}) and (\ref{eq13a}) along the real, positive
frequency axis follows from the identities \cite{59}
\begin{eqnarray}
&&
{\rm arctanh}x=\frac{1}{2}\ln\frac{1+x}{1-x},\quad 0\leq x^2<1,
\nonumber \\
&&
{\rm arctanh}x={\rm arctanh}\frac{1}{x}+i\frac{\pi}{2},
\label{eq13b}
\end{eqnarray}
\noindent
where the last identity is used for $x>1$ at the upper bank of the cut which
passes from unity to $\infty$.

The analytic properties of the functions
$\sigma(\omega,T)$, $\sigma^{(0)}(\omega)$ and
$\sigma^{(1)}(\omega,T)$ result in the validity of the Kramers-Kronig
relations which can be proven in exactly the same manner as it is done
in Ref.~\cite{55} for the case of dielectric permittivity.
The form of the Kramers-Kronig relations depends on the behavior of
$\sigma$ at zero frequency. As is seen in Eq.~(\ref{eq8}), both the real
and imaginary parts of $\sigma^{(0)}$ are regular at $\omega=0$
(the first order pole in the first term of ${\rm Im}\sigma^{(0)}$
is canceled by a similar pole with an opposite sign in the second term).
At $\omega\to\infty$ the quantity ${\rm Re}\sigma^{(0)}$ goes to $\sigma_0$.
Because of this, the Kramers-Kronig relation is valid for the function
${\rm Re}\sigma^{(0)}-\sigma_0$. The result is similar to that presented
in Ref.~\cite{55} for the dielectric permittivity
\begin{eqnarray}
&&
{\rm Re}\sigma^{(0)}(\omega)=\sigma_0+\frac{2}{\pi}\dashint_{0}^{\infty}
\frac{\xi {\rm Im}\sigma^{(0)}(\xi)}{\xi^2-\omega^2}d\xi,
\nonumber \\
&&
{\rm Im}\sigma^{(0)}(\omega)=-\frac{2\omega}{\pi}\dashint_{0}^{\infty}
\frac{{\rm Re}\sigma^{(0)}(\xi)}{\xi^2-\omega^2}d\xi,
\label{eq14}
\end{eqnarray}
\noindent
where the crossed sign of integration means that the principal value of
the integral is taken. We note also that
\begin{equation}
\dashint_{0}^{\infty}
\frac{d\xi}{\xi^2-\omega^2}=0.
\label{eq15}
\end{equation}
\noindent
Because of this it is not necessary to subtract $\sigma_0$ in the nominator
of the second equality in Eq.~(\ref{eq14}).

Now we consider the second contribution to the conductivity of graphene, i.e.,
$\sigma^{(1)}$. As is seen in Eq.~(\ref{eq10a}), the imaginary part of
$\sigma^{(1)}$ has the first-order pole. Because of this, the Kramers-Kronig
relations are similar to those obtained in Ref.~\cite{55} for the dielectric
permittivity of conductors
\begin{eqnarray}
&&
{\rm Re}\sigma^{(1)}(\omega,T)=\frac{2}{\pi}\dashint_{0}^{\infty}
\frac{\xi {\rm Im}\sigma^{(1)}(\xi,T)-C(T)}{\xi^2-\omega^2}d\xi,
\nonumber \\[-1mm]
&&
\label{eq16}\\[2mm]
&&
{\rm Im}\sigma^{(1)}(\omega,T)=-\frac{2\omega}{\pi}\dashint_{0}^{\infty}
\frac{{\rm Re}\sigma^{(1)}(\xi,T)}{\xi^2-\omega^2}d\xi
+\frac{C(T)}{\omega},
\nonumber
\end{eqnarray}
\noindent
where $C(T)$ is defined in Eq.~(\ref{eq10}).
We note that both the real and imaginary parts of  $\sigma^{(1)}$ defined
in Eq.~(\ref{eq10a}) go to zero when $\omega\to\infty$. Because of this, it is
not needed to subtract any constant from ${\rm Re}\sigma^{(1)}$ like it was
done in Eq.~(\ref{eq14}). At the same time, it is necessary to subtract
$C(T)$ in the nominator of the first dispersion relation in Eq.~(\ref{eq16}). This subtraction does not change the value of the integral at all $\omega\neq 0$
due to Eq.~(\ref{eq15}), but makes the Kramers-Kronig relation  correct at
$\omega=0$ (see the relevant discussions in Ref.~\cite{55} for the dielectric
permittivity of metals and in Sec.~III).

By combining Eqs.~(\ref{eq14}) and (\ref{eq16}), one arrives to the
Kramers-Kronig relations for the total conductivity of graphene at any
temperature
\begin{eqnarray}
&&
{\rm Re}\sigma(\omega,T)=\sigma_0+\frac{2}{\pi}\dashint_{0}^{\infty}
\frac{\xi {\rm Im}\sigma(\xi,T)-C(T)}{\xi^2-\omega^2}d\xi,
\nonumber \\[-1mm]
&&
\label{eq17}\\[2mm]
&&
{\rm Im}\sigma(\omega,T)=-\frac{2\omega}{\pi}\dashint_{0}^{\infty}
\frac{{\rm Re}\sigma(\xi,T)}{\xi^2-\omega^2}d\xi
+\frac{C(T)}{\omega}.
\nonumber
\end{eqnarray}

The Kramers-Kronig relations (\ref{eq14}), (\ref{eq16}), and (\ref{eq17})
follow from the discussed above general analytic properties of the local
conductivity of graphene. None of the expressions for the graphene
conductivity obtained in the previous literature using various approximate
and phenomenological methods satisfy these relations precisely.
Below we demonstrate, however, that the conductivity (\ref{eq7}), (\ref{eq8}),
(\ref{eq10a}), derived independently on the basis of first principles of
quantum electrodynamics at nonzero temperature using the polarization tensor,
is in full agreement with the Kramers-Kronig relations and, thus, with the
demands of causality.

\section{Kramers-Kronig relations for the conductivity at zero temperature}

We begin with the case $\Delta\geq 2\mu$ when the total conductivity of graphene
at $T=0$ is given by Eq.~(\ref{eq8}), i.e.,
$\sigma(\omega,0)=\sigma^{(0)}(\omega)$.
It is straightforward to substitute the first line of Eq.~(\ref{eq8})
in the right-hand side of the second Kramers-Kronig relation in
Eq.~(\ref{eq14}) and obtain
\begin{eqnarray}
&&
-\frac{2\omega}{\pi}\dashint_{0}^{\infty}
\frac{{\rm Re}\sigma^{(0)}(\xi)}{\xi^2-\omega^2}d\xi
\label{eq18}\\
&&
~=
-\frac{2\sigma_0}{\pi}\hbar\omega\left[\dashint_{\Delta}^{\infty}
\frac{d\zeta}{\zeta^2-\hbar^2\omega^2}+
\Delta^2\dashint_{\Delta}^{\infty}
\frac{d\zeta}{\zeta^2(\zeta^2-\hbar^2\omega^2)}\right],
\nonumber
\end{eqnarray}
\noindent
where the integration variable $\zeta=\hbar\xi$ is introduced.
Integrating on the right-hand side of Eq.~(\ref{eq18}) we find
\begin{eqnarray}
&&
-\frac{2\omega}{\pi}\dashint_{0}^{\infty}
\frac{{\rm Re}\sigma^{(0)}(\xi)d\xi}{\xi^2-\omega^2}=
\frac{\sigma_0}{\pi}\left[
\vphantom{\left|\frac{\hbar\omega+\Delta}{\hbar\omega-\Delta}\right|}
\frac{2\Delta}{\hbar\omega}\right.
\label{eq19}\\
&&
~\left.
-\frac{(\hbar\omega)^2+\Delta^2}{(\hbar\omega)^2}
\ln\left|\frac{\hbar\omega+\Delta}{\hbar\omega-\Delta}\right|
\right]={\rm Im}\sigma^{(0)}(\omega)
\nonumber
\end{eqnarray}
\noindent
if to take into account Eq.~(\ref{eq8}). Thus, the second Kramers-Kronig
relation in Eq.~(\ref{eq14}) is really satisfied.

Now we substitute the second line of Eq.~(\ref{eq8})
in the right-hand side of the first Kramers-Kronig relation in
Eq.~(\ref{eq14}) and obtain
\begin{widetext}
\begin{equation}
\sigma_0+\frac{2}{\pi}\dashint_{0}^{\infty}
\frac{\xi {\rm Im}\sigma^{(0)}(\xi)}{\xi^2-\omega^2}d\xi=
\sigma_0-\frac{2\sigma_0}{\pi^2}\dashint_{0}^{\infty}
\frac{\xi }{\xi^2-\omega^2}
\left[\ln\left|\frac{\hbar\xi+\Delta}{\hbar\xi-\Delta}\right|
+\frac{\Delta^2}{(\hbar\xi)^2}
\ln\left|\frac{\hbar\xi+\Delta}{\hbar\xi-\Delta}\right|\right]
d\xi,
\label{eq20}
\end{equation}
\end{widetext}
\noindent
where we have taken into account Eq.~(\ref{eq15}).

In the Appendix A, we calculate the following important integral:
\begin{equation}
I(b)\equiv\dashint_{0}^{\infty}\frac{y}{y^2-b^2}\ln\left|
\frac{y+1}{y-1}\right|dy=\left\{
\begin{array}{ll}
\frac{\pi^2}{2},& |b|<1, \\[1.5mm]
0, & |b|>1
\end{array}\right.
\label{eq21}
\end{equation}
\noindent
and indicate relevant incorrect results contained in Ref.~\cite{56}.

Introducing the variable $y=\hbar\xi/\Delta$ in the first integral on the
right-hand side of Eq.~(\ref{eq20}) and using Eq.~(\ref{eq21}),
one obtains
\begin{eqnarray}
&&
\dashint_{0}^{\infty}
\frac{\xi }{\xi^2-\omega^2}
\ln\left|\frac{\hbar\xi+\Delta}{\hbar\xi-\Delta}\right|d\xi=
\dashint_{0}^{\infty}
\frac{y}{y^2-b^2}\ln\left|
\frac{y+1}{y-1}\right|dy
\nonumber \\
&&~~
=\left\{
\begin{array}{ll}
\frac{\pi^2}{2},& \hbar\omega<\Delta, \\[1.5mm]
0, & \hbar\omega>\Delta, \quad b\equiv\frac{\hbar\omega}{\Delta}.
\end{array}\right.
\label{eq22}
\end{eqnarray}

The second integral on the right-hand side of Eq.~(\ref{eq20})
can be evaluated similarly
\begin{widetext}
\begin{eqnarray}
&&
\frac{\Delta^2}{\hbar^2}\dashint_{0}^{\infty}
\frac{1}{\xi (\xi^2-\omega^2)}
\ln\left|\frac{\hbar\xi+\Delta}{\hbar\xi-\Delta}\right|d\xi=
\dashint_{0}^{\infty}
\frac{1}{y(y^2-b^2)}\ln\left|
\frac{y+1}{y-1}\right|dy
\label{eq23}\\
&&~
=\frac{1}{b^2}\left[\dashint_{0}^{\infty}
\frac{y}{y^2-b^2}\ln\left|
\frac{y+1}{y-1}\right|dy -\dashint_{0}^{\infty}
\frac{dy}{y}\ln\left|
\frac{y+1}{y-1}\right|\right]
=\left\{
\begin{array}{ll}
0,& \hbar\omega<\Delta, \\[1.5mm]
-\frac{1}{2}\!\left(\frac{\pi\Delta}{\hbar\omega}\right)^2, & \hbar\omega>\Delta.
\end{array}\right.
\nonumber
\end{eqnarray}
\end{widetext}
\noindent
In obtaining this result we have used Eq.~(\ref{eq21}) for two times,
namely with $b\neq 0$ and $b=0$.

Substituting Eqs.~(\ref{eq22}) and (\ref{eq23}) in the right-hand side of
Eq.~(\ref{eq20}) and using the first line of Eq.~(\ref{eq8}), one arrives at
\begin{eqnarray}
&&
\sigma_0+\frac{2}{\pi}\dashint_{0}^{\infty}
\frac{\xi {\rm Im}\sigma^{(0)}(\xi)}{\xi^2-\omega^2}d\xi
=\left\{
\begin{array}{ll}
0,& \hbar\omega<\Delta, \\[1.5mm]
\sigma_0\frac{(\hbar\omega)^2+\Delta^2}{(\hbar\omega)^2}, & \hbar\omega>\Delta
\end{array}\right.
\nonumber \\
&&~~
={\rm Re}\sigma^{(0)}(\omega).
\label{eq24}
\end{eqnarray}
\noindent
Thus, the conductivity $\sigma^{(0)}$ in Eq.~(\ref{eq8}) satisfies the
first Kramers-Kronig relation in Eq.~(\ref{eq14}).

Now we continue to consider the case of zero temperature, but assume that
$\Delta<2\mu$. In this case it holds
\begin{equation}
\sigma^{(1)}(\omega,0)=\lim_{T\to 0}\sigma^{(1)}(\omega,T)\neq 0
\label{eq25}
\end{equation}
\noindent
and
\begin{equation}
\sigma(\omega,0)=\sigma^{(0)}(\omega)+
\sigma^{(1)}(\omega,0).
\label{eq26}
\end{equation}

Calculations show that under the condition $\Delta<2\mu$ we have \cite{51}
\begin{eqnarray}
&&
{\rm Re}\sigma(\omega,0)=\sigma_0\theta(\hbar\omega-2\mu)
\frac{(\hbar\omega)^2+\Delta^2}{(\hbar\omega)^2},
\label{eq27} \\
&&
{\rm Im}\sigma(\omega,0)=\frac{\sigma_0}{\pi}\left[
\frac{4\mu}{\hbar\omega}-
\frac{(\hbar\omega)^2+\Delta^2}{(\hbar\omega)^2}
\ln\left|\frac{\hbar\omega+2\mu}{\hbar\omega-2\mu}\right|
\right].
\nonumber
\end{eqnarray}
\noindent
Similar to Eq.~(\ref{eq8}), this result is valid at the real, positive
frequency axis. It is easily seen that in the limiting case $\omega\to 0$
one has
\begin{equation}
{\rm Im}\sigma(\omega,0)=\frac{C(0)}{\omega}+O\left(\frac{\hbar\omega}{2\mu}\right),
\label{eq28}
\end{equation}
\noindent
where
\begin{equation}
C(0)=\frac{\sigma_0}{\pi}\frac{(2\mu)^2-\Delta^2}{\hbar\mu}.
\label{eq29}
\end{equation}
\noindent
The last equation is also obtainable
as a particular case of Eq.~(\ref{eq10}) if one
puts there $T=0$. In so doing it is necessary to take into account that
at $T\to 0$ only the interval of $v$ from unity to $2\mu/\Delta$
contributes to the integral.

Taking into account that ${\rm Im}\sigma(\omega,0)$  has a pole at zero frequency,
the Kramers-Kronig relations are given in this case by Eq.~(\ref{eq17}) where
one should replace $\sigma(\omega,T)$ with $\sigma(\omega,0)$ and $C(T)$ with
$C(0)$. It is easily seen that both of them are satisfied. Really, substituting
the first line of Eq.~(\ref{eq27}) in the right-hand side of the second
Kramers-Kronig relation in Eq.~(\ref{eq17}) with $C$ defined in Eq.~(\ref{eq29})
and introducing the variable $\zeta=\hbar\xi$, one obtains
\begin{eqnarray}
&&
-\frac{2\omega}{\pi}\dashint_{0}^{\infty}
\frac{{\rm Re}\sigma(\xi,0)}{\xi^2-\omega^2}d\xi
+\frac{\sigma_0}{\pi}\frac{(2\mu)^2-\Delta^2}{\hbar\mu\omega}
\nonumber\\
&&
~=
-\frac{2\sigma_0}{\pi}\hbar\omega\left[\dashint_{2\mu}^{\infty}
\!\!\!\frac{d\zeta}{\zeta^2-\hbar^2\omega^2}+
\Delta^2\dashint_{2\mu}^{\infty}\!\!\!
\frac{d\zeta}{\zeta^2(\zeta^2-\hbar^2\omega^2)}\right],
\nonumber \\
&&~~~~
+\frac{\sigma_0}{\pi}\frac{(2\mu)^2-\Delta^2}{\hbar\mu\omega}.
\label{eq30}
\end{eqnarray}

Calculating the integrals in  Eq.~(\ref{eq30}), we arrive at
\begin{eqnarray}
&&
-\frac{2\omega}{\pi}\dashint_{0}^{\infty}
\frac{{\rm Re}\sigma(\xi,0)d\xi}{\xi^2-\omega^2}
+\frac{\sigma_0}{\pi}\frac{(2\mu)^2-\Delta^2}{\hbar\mu\omega}.
\label{eq31}\\
&&
=
\frac{\sigma_0}{\pi}\left[
\vphantom{\left|\frac{\hbar\omega+\Delta}{\hbar\omega-\Delta}\right|}
\frac{4\mu}{\hbar\omega}
-\frac{(\hbar\omega)^2+\Delta^2}{(\hbar\omega)^2}
\ln\left|\frac{\hbar\omega+2\mu}{\hbar\omega-2\mu}\right|
\right]={\rm Im}\sigma(\omega,0)
\nonumber
\end{eqnarray}
\noindent
in accordance with Eq.~(\ref{eq17}).

Now we verify the first Kramers-Kronig relation in Eq.~(\ref{eq17}),
when $C(T)$ is replaced with $C(0)$ from Eq.~(\ref{eq29}). It is more
illustrative to consider first the case $\omega\neq 0$ when $C(0)$
can be simply omitted due to Eq.~(\ref{eq15}).
Substituting
the second line of Eq.~(\ref{eq27}) in the right-hand side of the first
Kramers-Kronig relation in Eq.~(\ref{eq17}), we find
\begin{widetext}
\begin{equation}
\sigma_0+\frac{2}{\pi}\dashint_{0}^{\infty}
\frac{\xi {\rm Im}\sigma(\xi,0)}{\xi^2-\omega^2}d\xi=
\sigma_0-\frac{2\sigma_0}{\pi^2}\dashint_{0}^{\infty}
\frac{\xi }{\xi^2-\omega^2}
\left[\ln\left|\frac{\hbar\xi+2\mu}{\hbar\xi-2\mu}\right|
+\frac{\Delta^2}{(\hbar\xi)^2}
\ln\left|\frac{\hbar\xi+2\mu}{\hbar\xi-2\mu}\right|\right]
d\xi.
\label{eq32}
\end{equation}
\end{widetext}
\noindent
The first integral on the right-hand side of this equation is calculated like in
Eq.~(\ref{eq22}) with $y=\hbar\xi/(2\mu)$ and $b=\hbar\omega/(2\mu)$ using
Eq.~(\ref{eq21}). The result is given by Eq.~(\ref{eq22}) where $\Delta$ is
replaced with $2\mu$. The second integral is calculated like in Eq.~(\ref{eq23}).
It is equal to zero when $\hbar\omega<2\mu$ and to
$-(\pi\Delta)^2/(2\hbar^2\omega^2)$ when $\hbar\omega>2\mu$.
Substituting the values of both integrals in Eq.~(\ref{eq32}) and taking into
account the first line in Eq.~(\ref{eq27}), one finds
\begin{eqnarray}
&&
\sigma_0+\frac{2}{\pi}\dashint_{0}^{\infty}
\frac{\xi {\rm Im}\sigma(\xi,0)}{\xi^2-\omega^2}d\xi
=\left\{
\begin{array}{ll}
0,& \hbar\omega<2\mu, \\[1.5mm]
\sigma_0\frac{(\hbar\omega)^2+\Delta^2}{(\hbar\omega)^2}, & \hbar\omega>2\mu.
\end{array}\right.
\nonumber \\
&&~~
={\rm Re}\sigma(\omega,0).
\label{eq33}
\end{eqnarray}
\noindent
Thus, with account of Eq.~(\ref{eq15}), the first Kramers-Kronig relation
in Eq.~(\ref{eq17}) is proven for all $\omega\neq 0$.

At $\omega=0$ the validity of this Kramers-Kronig relation is achieved by
the subtraction of $C(0)$ in the first line of Eq.~(\ref{eq32}).
To see this, we substitute ${\rm Im}\sigma(\omega,0)$ from Eq.~(\ref{eq27})
and $C(0)$ from Eq.~(\ref{eq29}) in the right-hand side of the first
Kramers-Kronig relation of Eq.~(\ref{eq17}) at $\omega=0$ and obtain
\begin{widetext}
\begin{equation}
\sigma_0+\frac{2}{\pi}\dashint_{0}^{\infty}
\frac{\xi {\rm Im}\sigma(\xi,0)-C(0)}{\xi^2}d\xi=
\sigma_0+\frac{2\sigma_0}{\pi^2}\dashint_{0}^{\infty}
\!\!\!d\xi
\left[\frac{\Delta^2}{\hbar\mu\xi^2}-\frac{1}{\xi}
\ln\left|\frac{\hbar\xi+2\mu}{\hbar\xi-2\mu}\right|
-\frac{\Delta^2}{(\hbar\xi)^3}
\ln\left|\frac{\hbar\xi+2\mu}{\hbar\xi-2\mu}\right|\right]
.
\label{eq34}
\end{equation}
\end{widetext}
\noindent
Calculation of all the three integrals on the right-hand side of this
equation (see Appendix B) results in
\begin{equation}
\sigma_0+\frac{2}{\pi}\dashint_{0}^{\infty}
\frac{\xi {\rm Im}\sigma(\xi,0)-C(0)}{\xi^2}d\xi=
\sigma_0-\sigma_0=0,
\label{eq35}
\end{equation}
\noindent
as it should be because in accordance to the first line of Eq.~(\ref{eq27})
\begin{equation}
{\rm Re}\sigma(0,0)=0.
\label{eq36}
\end{equation}

This concludes the proof of the Kramers-Kroniog relations for the conductivity of
graphene at zero temperature and validates the fact that expressions (\ref{eq8})
for $\Delta>2\mu$ and (\ref{eq27}) for $\Delta<2\mu$ satisfy the condition
of causality. In some particular cases Eqs.~(\ref{eq8}) and (\ref{eq27}) have
been derived using various models and phenomenological approaches \
(see, for instance, Refs.~\cite{46,60,61}).
Note, however, that the additional terms in the conductivity of graphene at
zero temperature containing the $\delta$-function of $\omega$, which were obtained
within some approaches (see, e.g., Refs.~\cite{46,62}), are not obtainable in our
formalism based on the first principles of quantum electrodynamics.
Such terms would violate the Kramers-Kronig relations and, thus, lead to
contradiction with the principle of causality \cite{63}.

\section{Kramers-Kronig relations at nonzero temperature}

At first, we prove the validity of the Kramers-Kronig relations for the
temperature-dependent part of the conductivity of graphene
$\sigma^{(1)}(\omega,T)$ defined in Eq.~(\ref{eq10a}).
As usual, we start from the second Kramers-Kronig relation in Eq.~(\ref{eq16}).
Substituting
the first line of Eq.~(\ref{eq10a}) in the right-hand side of the second
Kramers-Kronig relation in Eq.~(\ref{eq16}), one obtains
\begin{eqnarray}
&&
-\frac{2\omega}{\pi}\dashint_{0}^{\infty}
\frac{{\rm Re}\sigma^{(1)}(\xi,T)}{\xi^2-\omega^2}d\xi
+\frac{C(T)}{\omega}
\label{eq37}\\
&&
~=
\frac{2\omega\sigma_0}{\pi\hbar^2}\dashint_{\frac{\Delta}{\hbar}}^{\infty}
\!\frac{(\hbar\xi)^2+\Delta^2}{\xi^2(\xi^2-\omega^2)}
F(\xi,T)d\xi+\frac{C(T)}{\omega}.
\nonumber
\end{eqnarray}
\noindent
Introducing the new integration variable $v=\hbar\xi/\Delta$
and using the secons line of  Eq.~(\ref{eq10a}), we find
\begin{eqnarray}
&&
-\frac{2\omega}{\pi}\dashint_{0}^{\infty}
\frac{{\rm Re}\sigma^{(1)}(\xi,T)}{\xi^2-\omega^2}d\xi
+\frac{C(T)}{\omega}
\label{eq38}\\
&&
=
\frac{2\sigma_0}{\pi}\hbar\omega\Delta\dashint_{1}^{\infty}
\!\!\!dv\frac{v^2+1}{v^2[(v\Delta)^2-(\hbar\omega)^2]}
F\left(\frac{v\Delta}{\hbar},T\right)+\frac{C(T)}{\omega}
\nonumber\\
&&
={\rm Im}\sigma^{(1)}(\omega,T),
\nonumber
\end{eqnarray}
\noindent
i.e., the second Kramers-Kronig relation in  Eq.~(\ref{eq16})
is satisfied.

Now we substitute
the second line of Eq.~(\ref{eq10a}) to the right-hand side of the first
Kramers-Kronig relation in Eq.~(\ref{eq16}). Taking into account that in
the second line of Eq.~(\ref{eq10a}) the pole term is already separated,
one can consider both cases $\omega\neq 0$ and $\omega=0$ simultaneously.
The result is
\begin{eqnarray}
&&
J\equiv\frac{2}{\pi}\dashint_{0}^{\infty}
\frac{\xi{\rm Im}\sigma^{(1)}(\xi,T)-C(T)}{\xi^2-\omega^2}d\xi
\label{eq39}\\
&&
=\frac{4\sigma_0}{\pi^2}\hbar\Delta\dashint_{0}^{\infty}
\frac{\xi^2d\xi}{\xi^2-\omega^2}\dashint_{1}^{\infty}
\!\!\!dv\frac{(v^2+1)
F\left(\frac{v\Delta}{\hbar},T\right)}{v^2[(v\Delta)^2-(\hbar\omega)^2]}
\nonumber \\
&&
=\frac{4\sigma_0}{\pi^2}\dashint_{0}^{\infty}
\frac{y^2dy}{y^2-b^2}\dashint_{1}^{\infty}
\!\!\!dv\frac{v^2+1}{v^2}
\frac{F\left(\frac{v\Delta}{\hbar},T\right)}{v^2-y^2},
\nonumber
\end{eqnarray}
\noindent
where the integration variable $y=\hbar\xi/\Delta$ was introduced
and $b=\hbar\omega/\Delta$.

Note that if $b<1$, i.e., $\hbar\omega<\Delta$, then $v\neq b$ holds
over the entire integration region from unity to infinity.
Taking into account that
\begin{equation}
\dashint_{0}^{\infty}\frac{y^2dy}{(y^2-b^2)(y^2-v^2)}=0
\quad\mbox{for}{\ }b\neq v,
\label{eq40}
\end{equation}
\noindent
one immediately concludes that $J=0$.

It remains to consider the case $b>1$, i.e., $\hbar\omega>\Delta$.
To deal with this case, we present our integral (\ref{eq39}) in the form
\begin{equation}
J=-\frac{2\sigma_0}{\pi^2}\dashint_{0}^{\infty}\!\!\!\!
\frac{y\,dy}{y^2-b^2}\dashint_{1}^{\infty}
\!\frac{v^2+1}{v^2}
F\left(\frac{v\Delta}{\hbar},T\right)
d\ln\left|\frac{v+y}{v-y}\right|.
\label{eq41}
\end{equation}

Integrating here by parts we find
\begin{eqnarray}
&&
J=-\frac{2\sigma_0}{\pi^2}\dashint_{0}^{\infty}\!\!\!\!
\frac{y\,dy}{y^2-b^2}
\label{eq42}\\
&&~~\times
\left\{\left.\left[\frac{v^2+1}{v^2}
F\left(\frac{v\Delta}{\hbar},T\right)
\ln\left|\frac{v+y}{v-y}\right|\right]\right|_{1}^{\infty}\right.
\nonumber \\
&&~~~~
\left.
-\dashint_{1}^{\infty}\ln\left|\frac{v+y}{v-y}\right|
d\left[\frac{v^2+1}{v^2}
F\left(\frac{v\Delta}{\hbar},T\right)\right]\right\}.
\nonumber
\end{eqnarray}
\noindent
Taking into account that in accordance to Eq.~(\ref{eq9a})
$F(x,T)\to 0$ when $x\to \infty$, Eq.~(\ref{eq42}) leads to
\begin{eqnarray}
&&
J=\frac{4\sigma_0}{\pi^2}F\left(\frac{\Delta}{\hbar},T\right)
\dashint_{0}^{\infty}\!\!\!\!
\frac{y\,dy}{y^2-b^2}\ln\left|\frac{1+y}{1-y}\right|
\label{eq43}\\
&&
+\frac{2\sigma_0}{\pi^2}\dashint_{0}^{\infty}\!\!\!\!
\frac{y\,dy}{y^2-b^2}
\dashint_{1}^{\infty}\!\!\ln\left|\frac{v+y}{v-y}\right|
d\left[\frac{v^2+1}{v^2}
F\left(\frac{v\Delta}{\hbar},T\right)\right].
\nonumber
\end{eqnarray}

The first integral on the right-hand side of this equation is equal to zero
due to Eq.~(\ref{eq21}) and, changing the integration order with respect to
$y$ and $v$, we have
\begin{equation}
J=\frac{2\sigma_0}{\pi^2}\dashint_{1}^{\infty}\!\!
d\left[\frac{v^2+1}{v^2}
F\left(\frac{v\Delta}{\hbar},T\right)\right]
\dashint_{0}^{\infty}\!\!\!\!
\frac{y\,dy}{y^2-b^2}
\ln\left|\frac{v+y}{v-y}\right|.
\label{eq44}
\end{equation}
\noindent
Now we introduce the integration variable $t=y/v$ in the last integral
and obtain
\begin{equation}
J=\frac{2\sigma_0}{\pi^2}\dashint_{1}^{\infty}\!\!
d\left[\frac{v^2+1}{v^2}
F\left(\frac{v\Delta}{\hbar},T\right)\right]
\dashint_{0}^{\infty}\!\!\!\!
\frac{t\,dt}{t^2-\tilde{b}^2}
\ln\left|\frac{t+1}{t-1}\right|,
\label{eq45}
\end{equation}
\noindent
where $\tilde{b}=b/v$ can be both larger and less than unity.
According to Eq.~(\ref{eq21}), the last integral on the right-hand side
of Eq.~(\ref{eq45}) is equal to zero if $\tilde{b}>1$ (i.e., $v<b$) and
to $\pi^2/2$  if $\tilde{b}<1$ (i.e., $v>b$).
As a result, Eq.~(\ref{eq45}) is simplified to
\begin{eqnarray}
&&
J=\sigma_0\dashint_{b}^{\infty}\!\!
d\left[\frac{v^2+1}{v^2}
F\left(\frac{v\Delta}{\hbar},T\right)\right]
\label{eq46} \\
&&~
=
-\sigma_0\frac{b^2+1}{b^2}F\left(\frac{b\Delta}{\hbar},T\right)
=-\sigma_0\frac{(\hbar\omega)^2+\Delta^2}{(\hbar\omega)^2}
F(\omega,T).
\nonumber
\end{eqnarray}

Combining together the results for $b<1$ (i.e., $\hbar\omega<\Delta$) and
$b>1$ (i.e., $\hbar\omega>\Delta$) and using the first line of Eq.~(\ref{eq9}),
we conclude from Eq.~(\ref{eq39}) that
\begin{equation}
J=-\sigma_0\theta(\hbar\omega-\Delta)
\frac{(\hbar\omega)^2+\Delta^2}{(\hbar\omega)^2}F(\omega,T)
={\rm Re}\sigma^{(1})(\omega,T),
\label{eq47}
\end{equation}
\noindent
i.e., the first Kramers-Kronig relation in Eq.~(\ref{eq16}) is satisfied.

The total conductivity of graphene at nonzero temperature is given by
Eq.~(\ref{eq7}). Using Eqs.~(\ref{eq8}) and (\ref{eq10a}), one obtains
\begin{widetext}
\begin{eqnarray}
&&
{\rm Re}\sigma(\omega,T)=\sigma_0\theta(\hbar\omega-\Delta)
\frac{(\hbar\omega)^2+\Delta^2}{(\hbar\omega)^2}
[1-F(\omega,T)],
\label{eq48} \\
&&
{\rm Im}\sigma(\omega,T)=\frac{\sigma_0}{\pi}\left\{
\left[\frac{2\Delta}{\hbar}+\frac{\pi C(T)}{\sigma_0}\right]
\frac{1}{\omega}-\frac{(\hbar\omega)^2+\Delta^2}{(\hbar\omega)^2}
\ln\left|\frac{\hbar\omega+\Delta}{\hbar\omega-\Delta}\right|
\right.
\nonumber \\
&&~~~~
\left.
+2\hbar\omega\Delta\dashint_{1}^{\infty}\!\!\!dv
\frac{v^2+1}{v^2[(v\Delta)^2-(\hbar\omega)^2]}
F\left(\frac{v\Delta}{\hbar},T\right)\right\}.
\nonumber
\end{eqnarray}
\end{widetext}

Note that ${\rm Re}\sigma(\omega,T)$ can be rewritten in especially simple
and transparent equivalent form. For this purpose we use the definition of $F$
in Eq.~(\ref{eq9a}) and the following identity
\begin{equation}
\frac{1}{2}-\frac{1}{e^y+1}=\frac{1}{2}\tanh\frac{y}{2}.
\label{eq49}
\end{equation}
\noindent
The result is
\begin{eqnarray}
&&
{\rm Re}\sigma(\omega,T)=\sigma_0\theta(\hbar\omega-\Delta)
\frac{(\hbar\omega)^2+\Delta^2}{2(\hbar\omega)^2}
\label{eq50} \\
&&~~~~~
\times
\left(\tanh\frac{\hbar\omega+2\mu}{4k_BT}+
\tanh\frac{\hbar\omega-2\mu}{4k_BT}\right).
\nonumber
\end{eqnarray}
\noindent
As to ${\rm Im}\sigma(\omega,T)$, simple asymptotic expressions for it
in different regions of parameters and the results of numerical computations
can be found in Refs.~\cite{49,50,51}.
Although in the general case of gapped graphene with nonzero chemical
potential Eqs.~(\ref{eq48}) and (\ref{eq50}) were derived in Ref.~\cite{51},
in different special cases similar dependences have been obtained previously
using various approaches based on the Kubo formalism and two-dimensional
Drude model (see, e.g., Refs.~\cite{64,65,66,67,68,69}).

The Kramers-Kronig relations (\ref{eq17}) for the total conductivity of
graphene (\ref{eq48}) are satisfied automatically, because they are obtained
by the combination of already proven Kramers-Kronig relations (\ref{eq14}) and
(\ref{eq16}) satisfied for $\sigma^{(0)}(\omega)$ and $\sigma^{(1)}(\omega,T)$,
respectively.

\section{Conclusions and discussion}

In the foregoing, we have investigated the problem of causality for the
conductivity of graphene in the framework of the Dirac model.
Until recently, only some partial results for the conductivity of graphene have
been obtained using some models and phenomenological approaches.
To investigate the problem of causality, we use the complete results for
the spatially local conductivity found on the basis of first principles of
thermal quantum field theory using the polarization tensor of graphene in
(2+1)-dimensional space-time \cite{49,50,51}. The spatially nonlocal corrections to
these results were shown to be of the order of $10^{-5}$ of the local
contributions and, thus, are of no physical significance in the framework
of Dirac's model.

General discussion of causality presented in the paper
leads to the conclusion that both the total conductivity
of graphene and contributions to it $\sigma^{(0)}(\omega)$,
depending on the band gap, and  $\sigma^{(1)}(\omega,T)$,
depending on the band gap and chemical potential, are
the analytic functions in the upper half-plane of
complex frequencies and possess all the standard
symmetry properties. Hence it follows that the real and
imaginary parts of the conductivity of graphene derived
in any specific formalism must satisfy the
Kramers-Kronig relations. The form of these relations,
as shown above, depends on the
presence of a pole at zero frequency and takes into account
an existence of the universal conductivity. There is no
pole for the conductivity of graphene at zero temperature
under the condition that the band gap is larger than twice
the chemical potential, and there is such a pole in all
remaining cases. The fulfilment of the Kramers-Kronig
relations can be considered as a basic guideline in
deciding which specific expression for the conductivity of
graphene is correct.

We have shown through the direct analytic calculations
that the real and imaginary parts of the conductivity of
graphene, found in Ref.~\cite{51} in the most general
case of nonzero temperature, band gap and chemical
potential on the basis of first principles of thermal
quantum field theory, satisfy both Kramers-Kronig
relations precisely. In the process, the values of two
important integrals in the widely used tables have been
corrected, which might be useful in the context of
dispersion relations for the scattering amplitudes in
quantum field theory. One can conclude that the obtained
results are not of only fundamental theoretical character,
but they also open fresh opportunities for the use of
Kramers-Kronig relations in different fields of physics
and for the interpretation of experimental data.

\section*{Acknowledgments}
The work of V.M.M.\ was partially supported by the Russian
Government
Program of Competitive Growth of Kazan Federal University.
\appendix*
\section{A}
\setcounter{equation}{0}
\renewcommand{\theequation}{A\arabic{equation}}

Here, we calculate the integral (\ref{eq21}) and correct relevant integrals
in Ref.~\cite{56} which are important for various applications in a wide context
of dispersion relations in different branches of physics.

The integral in Eq.~(\ref{eq21}) can be presented in the form
\begin{equation}
I(b)=\frac{1}{2}[I_{+}(b)+I_{-}(b)],
\label{A1}
\end{equation}
\noindent
where
\begin{eqnarray}
&&
I_{+}(b)=\dashint_0^{\infty}\!\!\frac{dy}{y+b}\ln\left|\frac{y+1}{y-1}\right|,
\nonumber \\
&&
I_{-}(b)=\dashint_0^{\infty}\!\!\frac{dy}{y-b}\ln\left|\frac{y+1}{y-1}\right|.
\label{A2}
\end{eqnarray}
\noindent
We consider the case $b\geq 0$, $b\neq 1$. It is easily seen that the integrals in
Eq.~(\ref{A2}) converge at the points $y=1$, $y=b$.
Integrating by parts in Eq.~(\ref{A2}), one obtains
\begin{equation}
I_{\pm}(b)=2\dashint_{0}^{\infty}\!\!\!dy\frac{\ln|y\pm b|}{y^2-1},
\label{A3}
\end{equation}
\noindent
where  the out-of-integral terms vanish and the lower indices $\pm$
 correspond to plus and minus on the right-hand side, respectively.

{}From Eq.~(\ref{A3}) we find the derivative of $I_{\pm}$ with respect to $b$
\begin{equation}
\frac{dI_{\pm}(b)}{db}=\pm2\dashint_{0}^{\infty}\!\!\!dy
\frac{1}{(y\pm b)(y^2-1)}.
\label{A4}
\end{equation}
\noindent
Calculating this integral, we obtain the result
\begin{equation}
\frac{dI_{\pm}(b)}{db}=\pm2\frac{\ln b}{1-b^2}.
\label{A5}
\end{equation}
\noindent
{}From this it follows that
\begin{equation}
\frac{dI(b)}{db}=\frac{1}{2}\left[
\frac{dI_{+}(b)}{db}+\frac{dI_{-}(b)}{db}\right]=0,
\label{A6}
\end{equation}
\noindent
i.e., $I(b)$ takes the constant values in the intervals [0,1) and
(1,$\infty$), where it is a continuous function.

Let us consider first the interval [0,1) and find the values
\begin{eqnarray}
&&
I_{+}(0)=I_{-}(0)=\dashint_{0}^{\infty}\!\frac{dy}{y}
\ln\left|\frac{y+1}{y-1}\right|
\nonumber \\
&&
~~=
\int_{0}^{1}\!\frac{dy}{y}
\ln\frac{1+y}{1-y}+\int_{1}^{\infty}\!\frac{dy}{y}
\ln\frac{y+1}{y-1}.
\label{A7}
\end{eqnarray}
\noindent
Changing the integration variable according to $y=1/x$ in the second
integral on the right-hand side of Eq.~(\ref{A7}), one obtains
\begin{equation}
I_{\pm}(0)=2\int_{0}^{1}\!\frac{dy}{y}
\ln\frac{1+y}{1-y}=
4\sum_{k=1}^{\infty}\frac{1}{2k-1}
\int_{0}^{1}\!\!\!y^{2k-2}dy.
\label{A8}
\end{equation}
\noindent
Calculating this integral and taking into account that \cite{70}
\begin{equation}
\sum_{k=1}^{\infty}\frac{1}{(2k-1)^2}
=\frac{\pi^2}{8},
\label{A9}
\end{equation}
\noindent
we find
\begin{equation}
I_{\pm}(0)=\frac{\pi^2}{2}.
\label{A10}
\end{equation}
\noindent
Then from Eq.~(\ref{A1}) it follows that $I(b)=\pi^2/2$ under the
condition $0\leq b<1$ in agreement with the first line of Eq.~(\ref{eq21}).

Note that it is also possible now to find the values of integrals
$I_{\pm}(b)$ at any $b$.
By integrating Eq.~(\ref{A5}) with respect to $b$ for $b<1$,
we have
\begin{equation}
I_{\pm}(b)=\pm\ln{b}\ln\frac{1+b}{1-b}\mp {\rm Li}_2(b)\pm {\rm Li}_2(-b)+
\frac{\pi^2}{2},
\label{A11}
\end{equation}
\noindent
where ${\rm Li}_n(x)$ is the polylogarithm function. This equation can be checked
by differentiation taking into account that
\begin{equation}
\frac{d{\rm Li}_2(\pm b)}{db}=-\frac{\ln(1\mp b)}{b}.
\label{A12}
\end{equation}
\noindent
The value of the arbitrary integration constant in Eq.~(\ref{A11}),
$C=\pi^2/2$, is determined from Eq.~(\ref{A10}) taking into account that
${\rm Li}_2(0)=0$.

The result (\ref{A11}) is in disagreement with the formula 2.6.14.27 of
Ref.~\cite{56} where the independent on $b$ value of the integrals
$I_{\pm}(b)$ equal to $\pi$ is indicated leading to an incorrect result
$I(b)=\pi$. This formula is also in contradiction with the formula
2.6.14.24. The latter is in agreement with our result (\ref{A10}).

Now we consider the case when $b$ varies in the interval ($1,\infty$),
where the dilogarithm function has a cut. Using Eq.~(\ref{A12}), one can
easily check that the integration of Eq.~(\ref{A5}) results in
\begin{equation}
I_{\pm}(b)=\pm\ln{b}\ln\frac{b+1}{b-1}\pm {\rm Li}_2\left(\frac{1}{b}\right)
\mp {\rm Li}_2\left(-\frac{1}{b}\right).
\label{A13}
\end{equation}
\noindent
The integration constant $C=0$ is found from the fact that $I_{\pm}(b)\to 0$
when $b\to\infty$. From Eqs.~(\ref{A1}) and (\ref{A13}) we have $I(b)=0$
over the entire interval ($1,\infty$) which concludes the proof of
Eq.~(\ref{eq21}).

The result (\ref{A13}) contradicts to the formula 2.6.14.26 of Ref.~\cite{56},
where instead of Eq.~(\ref{A13}) an incorrect value $I_{\pm}(b)=0$ is
indicated.

\appendix*
\section{B}
\setcounter{equation}{0}
\renewcommand{\theequation}{B\arabic{equation}}

Here, we calculate the integrals contained in Eq.~(\ref{eq34}).
Introducing the new variable $y=\hbar\xi/(2\mu)$, the right-hand side
of Eq.~(\ref{eq34}) takes the form
\begin{equation}
\sigma_0+\frac{2\sigma_0}{\pi^2}(I_1-I_2),
\label{B1}
\end{equation}
\noindent
where
\begin{eqnarray}
&&
I_1=\frac{\Delta^2}{(2\mu)^2}\dashint_{0}^{\infty}
\left(\frac{2}{y^2}-\frac{1}{y^3}\ln\left|
\frac{y+1}{y-1}\right|\right)dy,
\nonumber \\
&&
I_2=\dashint_{0}^{\infty}
\frac{dy}{y}\ln\left|
\frac{y+1}{y-1}\right|.
\label{B2}
\end{eqnarray}

Using Eq.~(\ref{eq21}) with $b=0$, which is proven in Appendix A, one obtains
$I_2=\pi^2/2$. Because of this below we consider only $I_1$.
It is easily seen that this integral converges at $y=0$. Really, for $y<1$
it holds
\begin{eqnarray}
\ln\left|\frac{y+1}{y-1}\right|&=&\ln\frac{1+y}{1-y}=
2\sum_{k=1}^{\infty}\frac{y^{2k-1}}{2k-1}
\nonumber \\
&=&
2y+2\sum_{k=1}^{\infty}\frac{y^{2k+1}}{2k+1}.
\label{B3}
\end{eqnarray}
\noindent
In a similar way, for $y>1$ one obtains
\begin{eqnarray}
\ln\left|\frac{y+1}{y-1}\right|&=&\ln\frac{y+1}{y-1}=
\ln\frac{1+\frac{1}{y}}{1-\frac{1}{y}}
\nonumber \\
&=&
2\sum_{k=1}^{\infty}\frac{1}{(2k-1)y^{2k-1}}.
\label{B4}
\end{eqnarray}

Substituting Eqs.~(\ref{B3}) and (\ref{B4}) in the first line of Eq.~(\ref{B2}),
we find
\begin{widetext}
\begin{eqnarray}
&&
I_1=\frac{\Delta^2}{(2\mu)^2}\left[\int_{0}^{1}
\left(\frac{2}{y^2}-\frac{1}{y^3}\ln
\frac{1+y}{1-y}\right)dy+\int_{1}^{\infty}
\left(\frac{2}{y^2}-\frac{1}{y^3}\ln
\frac{y+1}{y-1}\right)dy\right]
\nonumber \\
&&
~~=
-\frac{2\Delta^2}{(2\mu)^2}\left\{\int_{0}^{1}
\sum_{k=1}^{\infty}\frac{y^{2k-2}}{2k+1}dy-
\int_{1}^{\infty}\left[\frac{1}{y^2}-
\sum_{k=1}^{\infty}\frac{1}{(2k-1)y^{2k+2}}\right]dy\right\}.
\label{B5}
\end{eqnarray}
\end{widetext}

Integrating on the right-hand side of this equation, one arrives at
\begin{equation}
I_1=-\frac{2\Delta^2}{(2\mu)^2}\left[
\sum_{k=1}^{\infty}\frac{1}{4k^2-1} -1+
\sum_{k=1}^{\infty}\frac{1}{4k^2-1}\right].
\label{B6}
\end{equation}
\noindent
Taking into account that \cite{70}
\begin{equation}
\sum_{k=1}^{\infty}\frac{1}{4k^2-1}=\frac{1}{2},
\label{B7}
\end{equation}
\noindent
we finally obtain that $I_1=0$. Substituting the values of both $I_1$
and $I_2$ in Eq.~(\ref{B1}), one finds
\begin{equation}
\sigma_0+\frac{2\sigma_0}{\pi^2}(I_1-I_2)=0
\label{B8}
\end{equation}
\noindent
in accordance with Eq.~(\ref{eq35}).


\begin{thebibliography}{99}
\bibitem{1}
M.~I.~Katsnelson,
{\it Graphene: Carbon in Two Dimensions}
(Cambridge University Press, Cambridge, 2012).
\bibitem{2}
{\it Physics of Graphene}, ed. H.\ Aoki and M.\ S.\ Dresselhaus
(Springer, Cham, 2014).
\bibitem{3}
A.~H.~Castro Neto, F.~Guinea, N.~M.~R.~Peres, K.~S.~Novoselov,
and A.~K.~Geim,
The electronic properties of graphene,
Rev. Mod. Phys. {\bf 81}, 109 (2009).
\bibitem{4}
M.~I.~Katsnelson, K.~S.~Novoselov, and A.~K.~Geim,
Chiral tunneling and the Klein paradox in graphene,
Nature Phys. {\bf 2}, 620
(2006).
\bibitem{5}
D.~Allor, T.~D.~Cohen, and D.~A.~McGady,
Schwinger mechanism and graphene,
Phys. Rev. D {\bf 78}, 096009 (2008).
\bibitem{6}
C.~G.~Beneventano, P.~Giacconi, E.~M.~Santangelo,
and R.~Soldati,
Planar QED at finite temperature and density: Hall conductivity,
Berry's phases and minimal conductivity of graphene,
J. Phys. A {\bf 42}, 275401 (2009).
\bibitem{7}
G.~L.~Klimchitskaya and V.~M.~Mostepanenko,
Creation of quasiparticles by a time-dependent electric field,
{Phys. Rev.} D {\bf 87}, 125011 (2013).
\bibitem{8}
I.~Akal, R.~Egger, C.~M\"{u}ller, and S.\ Villarba-Ch\'{a}vez,
Low-dimensional approach to pair production in an oscillating electric
field: Application to bandgap graphene layers,
{Phys. Rev.} D {\bf 93}, 116006 (2016).
\bibitem{9}
M.~O.~Goerbig,
Electronic properties of graphene in a strong magnetic field,
Rev. Mod. Phys. {\bf 83}, 1193 (2011).
\bibitem{10}
P.~K.~Pyatkovsky,
Dynamical polarization, screening, and plasmons in gapped graphene,
J. Phys.: Condens. Matter {\bf 21}, 025506 (2009).
\bibitem{11}
M.~Bordag, I.~V.~Fialkovsky, D.~M.~Gitman, and
D.~V.~Vassilevich,
Casimir interaction between a perfect conductor and graphene
described by the Dirac model,
{Phys. Rev. B} {\bf 80}, 245406 (2009).
\bibitem{12}
I.~V.~Fialkovsky, V.~N.~Marachevsky, and
D.~V.~Vassilevich,
Finite-temperature Casimir effect for graphene,
{Phys. Rev. B} {\bf 84}, 035446 (2011).
\bibitem{13}
M.~Bordag, G.~L.~Klimchitskaya, and
V.\ M.\ Mostepanenko,
Thermal Casimir effect in the interaction of graphene
with dielectrics and metals,
Phys. Rev. B {\bf 86}, 165429 (2012).
\bibitem{14}
M.~Chaichian, G.~L.~Klimchitskaya, V.\ M.\ Mostepanenko,
and A.~Tureanu,
Thermal Casimir-Polder interaction of different
atoms with graphene,
Phys. Rev. A {\bf 86}, 012515 (2012).
\bibitem{15}
G.~L.~Klimchitskaya
and V.~M.~Mostepanenko,
Van der Waals and Casimir interactions between
two graphene sheets,
{Phys. Rev.} B {\bf 87}, 075439 (2013).
\bibitem{16}
B.~Arora, H.~Kaur, and B.~K.~Sahoo,
$C_3$ coefficients for the alkali atoms interacting with a graphene
 and carbon nanotube,
J. Phys. B {\bf 47}, 155002 (2014).
\bibitem{17}
K.~Kaur, J.~Kaur, B.~Arora,  and B.~K.~Sahoo,
Emending thermal dispersion interaction of Li, Na, K and Rb
alkali-metal atoms with graphene in the Dirac model,
Phys. Rev. B {\bf 90}, 245405 (2014).
\bibitem{18}
G.~L.~Klimchitskaya and V.~M.~Mostepanenko,
Classical Casimir-Polder force between polarizable
microparticles and thin films including graphene,
{Phys. Rev.} A {\bf 89}, 012516 (2014).
\bibitem{19}
G.~L.~Klimchitskaya and V.~M.~Mostepanenko,
Classical limit of the Casimir interaction for thin films
with applications to graphene,
{Phys. Rev.} B {\bf 89}, 035407 (2014).
\bibitem{20}
G.~L.~Klimchitskaya and V.~M.~Mostepanenko,
Observability of thermal effects in the Casimir interaction
from graphene-coated substrates,
{Phys. Rev.} A {\bf 89}, 052512 (2014).
\bibitem{21}
G.~L.~Klimchitskaya and V.~M.~Mostepanenko,
Impact of graphene coating on the atom-plate interaction,
{Phys. Rev.} A {\bf 89}, 062508 (2014).
\bibitem{22}
G.~L.~Klimchitskaya, V.~M.~Mostepanenko, and
Bo~E.~Sernelius,
Two approaches for describing the Casimir interaction with graphene:
density-density correlation function versus polarization tensor,
Phys. Rev. B {\bf 89}, 125407 (2014).
\bibitem{23}
G.~L.~Klimchitskaya, U.~Mohideen, and V.~M.~Mostepanenko,
Theory of the Casimir interaction for graphene-coated substrates
 using the polarization tensor and comparison with experiment,
Phys. Rev. B {\bf 89}, 115419 (2014).
\bibitem{24}
J.~F.~Dobson, A.~White, and A.~Rubio,
Asymptotics of the dispersion interaction: Analytic
benchmarks for van der Waals energy functionals,
{Phys. Rev. Lett.} {\bf 96}, 073201 (2006).
\bibitem{25}
G.~G\'{o}mez-Santos,
Thermal van der Waals interaction between graphene layers,
Phys. Rev. B {\bf 80}, 245424 (2009).
\bibitem{26}
D.~Drosdoff and L.~M.~Woods,
Casimir forces and graphene sheets,
Phys. Rev. B {\bf 82}, 155459 (2010).
\bibitem{27}
D.~Drosdoff and L.~M.~Woods,
Casimir interaction between graphene sheets and metamaterials,
Phys. Rev. A {\bf 84}, 062501 (2011).
\bibitem{28}
Bo~E.~Sernelius,
Casimir interactions in graphene systems,
Europhys. Lett. {\bf 95}, 57003 (2011).
\bibitem{29}
T.~E.~Judd, R.~G.~Scott, A.~M.~Martin, B.\ Kaczmarek,
and T.\ M.\ Fromhold,
Quantum reflection of ultracold atoms from thin films, graphene
and semiconductor heterostructures,
New J. Phys. {\bf 13}, 083020 (2011).
\bibitem{30}
J.~Sarabadani, A.~Naji, R.~Asgari, and R.~Podgornik,
Many-body effects in the van der Waals-Casimir interaction
between graphene layers,
Phys. Rev. B {\bf 84}, 155407 (2011);
Phys. Rev. B {\bf 87}, 239905(E) (2013).
\bibitem {31}
D.~Drosdoff, A.~D.~Phan, L.~M.~Woods, I.\ V.\ Bondarev,
and J.\ F.\ Dobson,
Effects of spatial dispersion on the Casimir force between
graphene sheets,
Eur. Phys. J. B {\bf 85}, 365 (2012).
\bibitem{32}
Bo~E.~Sernelius,
Retarded interactions in graphene systems,
{Phys. Rev.} B {\bf 85}, 195427 (2012).
\bibitem{33}
A.~D.~Phan, L.~M.~Woods, D.~Drosdoff,
I.\ V.\ Bondarev, and N.\ A.\ Viet,
Temperature dependent graphene suspension due to thermal
Casimir interaction,
Appl. Phys. Lett. {\bf 101}, 113118 (2012).
\bibitem{34}
A.~D.~Phan, N.\ A.\ Viet, N.\ A.\ Poklonski, L.~M.~Woods,
 and  C.\ H.\ Le,
Interaction of a graphene sheet with a ferromagnetic metal plate,
Phys. Rev. B {\bf 86}, 155419 (2012).
\bibitem{35}
N.~Knusnutdinov, R.~Kashapov, and L.~M.~Woods,
Casimir-Polder effect for a stack of conductive planes,
 Phys. Rev. A {\bf 94}, 012513 (2016).
\bibitem{36}
M.~Bordag, G.~L.~Klimchitskaya, V.~M.~Mostepanenko, and V.~M.~Petrov,
Quantum field theoretical description for the reflectivity of graphene,
Phys. Rev. D {\bf 91}, 045037 (2015); {\bf 93}, 089907(E) (2016).
\bibitem{37}
G.~L.~Klimchitskaya and V.~M.~Mostepanenko,
Origin of large thermal effect in the Casimir interaction between
two graphene sheets,
{Phys. Rev.} B {\bf 91}, 174501 (2015).
\bibitem{38}
G.~L.~Klimchitskaya,
Quantum field theory of the Casimir force for graphene,
{Int. J. Mod. Phys.} A {\bf 31}, 1641026 (2016).
\bibitem{39}
V.~B.~Bezerra, G.~L.~Klimchitskaya,
V.~M.~Mostepanenko, and C.\ Romero,
Nernst heat theorem for the thermal Casimir interaction
between two graphene sheets,
{Phys. Rev.} A {\bf 94}, 042501 (2016).
\bibitem{40}
G.~Bimonte, G.~L.~Klimchitskaya, and V.~M.~Mostepanenko,
How to observe the giant thermal effect in the Casimir force
for graphene systems,
 Phys. Rev. A {\bf 96}, 012517 (2017).
\bibitem{40a}
C.~Henkel, G.~L.~Klimchitskaya, and V.~M.~Mostepanenko,
Influence of chemical potential on the Casimir-Polder interaction
between an atom and gapped graphene or graphene-coated substrate,
 Phys. Rev. A {\bf 97}, 032504 (2018).
\bibitem{41}
G.~L.~Klimchitskaya, C.~C.~Korikov, and V.~M.~Petrov,
Theory of reflectivity properties of graphene-coated material plates,
Phys. Rev. B {\bf 92}, 125419 (2015); {\bf 93}, 159906(E) (2016).
\bibitem{42}
G.~L.~Klimchitskaya and V.~M.~Mostepanenko,
Reflectivity properties of graphene with nonzero mass-gap parameter,
{Phys. Rev.} A {\bf 93}, 052106 (2016).
\bibitem{43}
G.~L.~Klimchitskaya and V.~M.~Mostepanenko,
Optical properties of dielectric plates coated with gapped graphene,
{Phys. Rev.} B {\bf 95}, 035425 (2017).
\bibitem{44}
M.~Bordag, I.~Fialkovskiy, and D.~Vassilevich,
Enhanced Casimir effect for doped graphene,
Phys. Rev. B {\bf 93}, 075414 (2016);
{\bf 95}, 119905(E) (2017).
\bibitem{45}
G.~Bimonte, G.~L.~Klimchitskaya, and V.~M.~Mostepanenko,
Thermal effect in the Casimir force for graphene and graphene-coated
substrates: Impact of nonzero mass gap and chemical potential,
 Phys. Rev. B {\bf 96}, 115430 (2017).
\bibitem{46}
V.~P.~Gusynin, S.~G.~Sharapov, and J.~P.~Carbotte,
AC conductivity of graphene: From tight-binding model to
2+1-dimensional quantum electrodynamics,
Int. J. Mod. Phys. B {\bf 21}, 4611 (2007).
\bibitem{47}
N.~M.~R.~Peres,
The transport properties of graphene: An introduction,
Rev. Mod. Phys. {\bf 82}, 2673 (2010).
\bibitem{48}
S.~Das~Sarma, S.~Adam, E.~H.~Hwang, and E.\ Rossi,
Electronic transport in two-dimensional graphene,
Rev. Mod. Phys. {\bf 83}, 407 (2011).
\bibitem{49}
G.~L.~Klimchitskaya and V.~M.~Mostepanenko,
Conductivity of pure graphene: Theoretical approach using the polarization tensor,
{Phys. Rev.} B {\bf 93}, 245419 (2016).
\bibitem{50}
G.~L.~Klimchitskaya and V.~M.~Mostepanenko,
Quantum electrodynamic approach to the conductivity of gapped graphene,
{Phys. Rev.} B {\bf 94}, 195405 (2016).
\bibitem{51}
G.~L.~Klimchitskaya, V.~M.~Mostepanenko, and V.~M.~Petrov,
Conductivity of graphene in the framework of Dirac model:
Interplay between nonzero mass gap and chemical potential,
{Phys. Rev.} B {\bf 96}, 235432 (2017).
\bibitem{52}
D.~Liu and S.~Zhang,
Kramers-Kronig relation of graphene conductivity,
J. Phys.: Condens. Matter {\bf 20}, 175222 (2008).
\bibitem{52a}
M.~Jablan, H.~Buljan, and M.~Solja\v{c}i\'{c},
Transverse electric plasmons in bilayer graphene,
Optics Express {\bf 19}, 11236 (2011).
\bibitem{53}
J.~Horng, Chi-Fan Chen, B.\ Geng, C.\ Girit, Y.\ Zhang, Z.\ Hao,
H.\ A.\ Bechtel, M.\ Martin, A.\ Zettl, M.\ F.\ Crommie,
Y.\ R.\ Shen, and F.\ Wang,
Drude conductivity of Dirac fermions in graphene,
{Phys. Rev.} B {\bf 83}, 165113 (2011).
\bibitem{54}
V.\ U.\ Nazarov,
Negative static permittivity and violation of Kramers-Kronig
relations in quasi-two-dimensional crystals,
{Phys. Rev.} B {\bf 92}, 161402(R) (2015).
\bibitem{55}
L.~D.~Landau, E.~M.~Lifshitz, and L.~P.~Pitaevskii,
{\it Electrodynamics of Continuous Media}
(Pergamon, Oxford, 1984).
\bibitem{56}
A.~P.~Prudnikov, Yu.~A.~Brychkov, and O.\ I.\ Marichev,
{\it Integrals and Series. Vol.1. Elementary Functions}
(Gordon and Breach, New York, 1986).
\bibitem{57}
A.~I.~Akhiezer and V.~B.~Berestetskii, {\it Quantum Electrodynamics} (Interscience,
New York, 1965).
\bibitem{58}
S.~S.~Schweber, {\it An Introduction to Relativistic Quantum Field Theory}
(Dover, New York, 2005).
\bibitem{59}
{\it Handbook of Mathematical Functions with Formulas, Graphs and
Mathematical Tables}, eds. M.\ Abramowitz and I.\ A.\ Stegun
(Dover Publications, New York, 2012).
\bibitem{60}
T.~G.~Pedersen, A.-P.~Jauho, and K.~Pedersen,
Optical response and excitons in gapped graphene,
Phys. Rev. B {\bf 79}, 113406 (2009).
\bibitem{61}
T.~Stauber,
Plasmonics in Dirac systems: from graphene to topological insulators,
J. Phys.: Condens. Matter {\bf 26}, 123201 (2014).
\bibitem{62}
V.~P.~Gusynin, S.~G.~Sharapov, and J.\ P.\ Carbotte,
Magneto-optical conductivity in graphene,
J. Phys.: Condens. Matter {\bf 19}, 026222 (2007).
\bibitem{63}
G.~L.~Klimchitskaya and V.~M.~Mostepanenko,
Comment on ``Lifshitz-Matsubara sum formula for the Casimir pressure
between magnetic metallic mirrors",
{Phys. Rev.} E {\bf 94}, 026101 (2016).
\bibitem{64}
S.~Ryu, C.~Mudry, A.~Furusaki, and A.\ W.\ W.\ Ludwig,
Landauer conductance and twisted boundary conditions for Dirac fermions
in two space dimensions,
{Phys. Rev.} B {\bf 75}, 205344 (2007).
\bibitem{65}
L.~A.~Falkovsky and S.~S.~Pershoguba,
Optical far-infrared properties of a graphene
monolayer and multilayer,
Phys. Rev. B {\bf 76}, 153410 (2007).
\bibitem{66}
L.~A.~Falkovsky and A.~A.~Varlamov,
Space-time dispersion of graphene conductivity,
Eur. Phys. J. B {\bf 56}, 281 (2007).
\bibitem{67}
T.~Stauber, N.~M.~R.~Peres, and A.~K.~Geim,
Optical conductivity of graphene in the visible region of the spectrum,
Phys. Rev. B {\bf 78}, 085432 (2008).
\bibitem{68}
K.~F.~Mak, M.~Y.~Sfeir, Y.~Wu, C.~H.~Lui, J.\ A.\ Misewich,
and T.\ F.\ Heinz,
Measurement of the optical conductivity of graphene,
Phys. Rev. Lett. {\bf 101}, 196405 (2008).
\bibitem{69}
L.~A.~Falkovsky,
Optical properties of graphene,
J. Phys.: Conf. Series {\bf 129}, 012004 (2008).
\bibitem{70}
I.~S.~Gradshtein and I.~M.~Ryzhik,
{\it Table of Integrals, Series and Products}
(Academic Press, New York, 1980).
\end{thebibliography}
\end{document}